\documentclass[aps,prl,twocolumn,preprintnumbers, showpacs, nofootinbib, superscriptaddress]{revtex4-1}
\usepackage[utf8]{inputenc}
\usepackage{graphicx}
\usepackage{amsmath,amssymb}
\usepackage{xcolor}

\usepackage[hypertexnames=false]{hyperref}
\hypersetup{
    colorlinks=true,       
    linkcolor=blue,          
    citecolor=blue,        
    filecolor=blue,      
    urlcolor=blue           
}

\newcommand{\obs}[0]{\mathcal{O}}
\newcommand{\mobs}[0]{\mathcal{Q}}
\newcommand{\mparam}[0]{\vec{\omega}}
\DeclareMathOperator{\Var}{Var}
\let\Im\undefined
\let\Re\undefined
\DeclareMathOperator{\Re}{Re}
\DeclareMathOperator{\Im}{Im}
\DeclareMathOperator{\grad}{\nabla}
\DeclareMathOperator{\arccosh}{arccosh}
\newcommand{\Seff}[0]{S_{\mathrm{eff}}}

\newcommand{\ssection}[1]{\textbf{#1} --- }

\begin{document}

\title{Path integral contour deformations for noisy observables}

\author{William Detmold }
\affiliation{Center for Theoretical Physics, Massachusetts Institute of Technology, Cambridge, MA 02139, USA}
\author{Gurtej Kanwar}
\affiliation{Center for Theoretical Physics, Massachusetts Institute of Technology, Cambridge, MA 02139, USA}
\author{Michael L. Wagman}
\affiliation{Center for Theoretical Physics, Massachusetts Institute of Technology, Cambridge, MA 02139, USA}
\affiliation{Fermi National Accelerator Laboratory, Batavia, IL 60510, USA}
\author{Neill C. Warrington}
\affiliation{Institute for Nuclear Theory, University of Washington, Seattle, Washington 98195-1550}
\preprint{FERMILAB-PUB-20-095-T}
\preprint{INT-PUB-20-007}
\preprint{MIT-CTP/5182}

\begin{abstract}
Monte Carlo studies of many quantum systems face exponentially severe signal-to-noise problems. We show that noise arising from complex phase fluctuations of observables can be reduced without introducing bias using path integral contour deformation techniques. A numerical study of contour deformations for correlation functions in Abelian gauge theory and complex scalar field theory demonstrates that variance can be reduced by orders of magnitude without modifying Monte Carlo sampling.
\end{abstract}

\maketitle

Understanding the dynamics of strongly coupled quantum systems is a fundamental challenge in many contexts including nuclear structure and reactions, condensed matter and cold atomic physics, and new physics searches using hadrons and nuclei as probes.
Strongly coupled quantum theories generically cannot be solved analytically, and Monte Carlo (MC) methods are typically used to calculate expectation values of observables in these theories by sampling over high-dimensional configuration spaces.

In lattice quantum field theories (QFTs) it has long been realized that signal-to-noise (StN) ratios of MC estimates of imaginary-time correlation functions are exponentially small in the separations between the defining operators~\cite{Parisi:1983ae,Lepage:1989hd}.
These \emph{signal-to-noise problems} become exponentially more severe for systems with increasing charge and, for example, limit lattice quantum chromodynamics (QCD) calculations of nuclei to systems with baryon number $A \leq 5$~\cite{Beane:2009kya,Beane:2009gs,Yamazaki:2009ua,Doi:2011gq,Doi:2012xd,Detmold:2012eu,Yamazaki:2012hi,Detmold:2014hla,Chang:2015qxa,Yamazaki:2015asa,Savage:2016kon,Winter:2017bfs,Chang:2017eiq} and obstruct calculations of quantities needed to interpret experiments seeking to identify new physics using large nuclei~\cite{Detmold:2019ghl,Kronfeld:2019nfb,Cirigliano:2019jig}.
Similar StN problems obstruct calculations in nuclear many-body theories~\cite{Wiringa:2000gb,Carlson:2014vla,Lahde:2015ona}, in spin and isospin asymmetric nuclear matter encountered in nuclear astrophysics~\cite{Gandolfi:2014ewa,PhysRevC.83.065801,Fantoni:2001ih}, and in quantum MC studies of non-relativistic fermions in condensed matter~\cite{Zhang:1995zz,Zhang:1996us,Endres:2011er,Endres:2012cw,Shi:2015lyu,Drut:2015uua,Porter:2016vry} and cold atomic physics \cite{Carlson:2005kg,Rammelmuller:2018hnk} contexts.

Correlation functions in imaginary time can be represented as path integrals of the form
\begin{equation}
   \begin{split}
      \left< \obs \right> \equiv \frac{1}{Z} \int_{\mathcal{M}} \mathcal{D} U \; e^{-S(U)} \; \obs(U),
   \end{split}\label{eq:obs}
\end{equation}
where $U$ is a quantum field, $\mathcal{M}$ is the integration manifold describing field configuration space, $\obs$ is a suitable product of fields that 
includes creation and annihilation operators for the quantum numbers of interest, and the partition function is
$      Z \equiv \int_{\mathcal{M}} \mathcal{D} U \; e^{-S(U)}.$
The action $S$ is assumed to be real. 
For baryon correlation functions in QCD, it was demonstrated in Ref.~\cite{Wagman:2016bam} that the StN problem arises from quantum fluctuations of the complex phase of $\mathcal{O}$.
A similar StN problem arises for correlation functions of charged scalar fields, where averaging over phase fluctuations is required to project correlation functions to particular charge sectors~\cite{Detmold:2018eqd}.
These complex phase fluctuations imply that the integrand of Eq.~\eqref{eq:obs} is not positive-definite or real and, as in systems with complex actions, the integral is determined by near cancellation of contributions with complex phases resulting in a \emph{sign problem}.

In certain cases, methods have been developed to exponentially improve sign and StN problems~\cite{Luscher:2001up,Meyer:2002cd,DellaMorte:2007zz,Ejiri:2007ga,DellaMorte:2008jd,DellaMorte:2010yp,Grabowska:2012ik,Langfeld:2012ah,Sexty:2014zya,Scorzato:2015qts,Langfeld:2015fua,Banuls:2016jws,Vera:2016xpp}. For example, in dual-variable approaches, integrals over phase fluctuations are computed analytically and sign problems are completely solved ~\cite{Ukawa:1979yv,PhysRevD.75.065012,2013slft.confE..38K,Gattringer:2012ap,Gattringer:2015nea,Gattringer:2016kco,GATTRINGER2018344,BRUCKMANN2015495}.
However, it remains an open challenge to extend these methods to generic observables in complicated QFTs such as QCD.
Other methods for taming StN problems such as phase unwrapping~\cite{Detmold:2018eqd} and multilevel integration for approximately factorizable correlation functions~\cite{Ce:2016ajy,Ce:2016idq,Ce:2019yds} can be applied to generic observables in complicated QFTs but introduce additional systematic uncertainties.

This letter introduces a general, exact method for improving the StN of noisy observables in theories with real actions. Noting that StN problems for baryon and other correlation functions arise from complex phase fluctuations~\cite{Wagman:2016bam},
we adapt manifold deformation techniques that have been used previously to address sign problems in QFTs with complex actions~\cite{Cristoforetti:2012su,Aarts:2013fpa,Cristoforetti:2013wha,Mukherjee:2013aga,Aarts:2014nxa,Cristoforetti:2014gsa,Alexandru:2015xva,Alexandru:2015sua,Alexandru:2016gsd,Fujii:2015vha,Tanizaki:2015rda,Alexandru:2017czx,Alexandru:2017lqr,Mori:2017nwj,Tanizaki:2017yow,Alexandru:2018brw,Alexandru:2018ngw,Alexandru:2018fqp,Alexandru:2018ddf,Kashiwa:2018vxr,Fukuma:2019uot,Fukuma:2019wbv,Kashiwa:2019lkv,Mou:2019gyl,Ulybyshev:2019fte} to correlation function StN problems.
Manifold deformation techniques are based on Cauchy's integral theorem, which states that integrals of holomorphic functions are unchanged when the domain of integration is smoothly deformed. Applied to path integrals, Cauchy's theorem implies that holomorphic observables, including correlation functions, are unchanged if the integration contour is deformed. However, the variance of a correlation function is non-holomorphic when phase fluctuations are present, and therefore will change.
If integration contours with lower variance can be found, then StN problems for observables can be reduced without changing their expectation values.
Methods for finding such contours are investigated in this work.

\ssection{Deformed observables}
Cauchy's theorem states that the integral of a holomorphic function is unchanged when the manifold of integration $\mathcal{M}$ is continuously deformed to manifold $\widetilde{\mathcal{M}}$, provided $\mathcal{M}$ can be deformed into $\widetilde{\mathcal{M}}$ without crossing non-analyticities of the integrand. Often the integrand of Eq.~\eqref{eq:obs}, $e^{-S}\obs$, may be analytically continued to a holomorphic function over complexified field space (see e.g.~Refs.~\cite{Aarts:2013uxa,Alexandru:2018ngw}), and therefore the domain of integration can be deformed without changing the path integral result. In this case, a manifold $\widetilde{\mathcal{M}}$ satisfying the requirements of Cauchy's theorem gives identical expectation values for $\obs$:
\begin{equation}
   \begin{split}
      \left< \obs \right> &= \frac{1}{Z} \int_{\widetilde{\mathcal{M}}} \mathcal{D} \widetilde{U} \; e^{-S(\widetilde{U})} \; \obs(\widetilde{U}) \\
      &= \frac{1}{Z} \int_{\mathcal{M}} \mathcal{D} U \; J(U)\; e^{-S(\widetilde{U}(U))} \; \obs(\widetilde{U}(U)).
   \end{split}\label{eq:def}
\end{equation}
Here $\widetilde{U}: \mathcal{M} \rightarrow \widetilde{\mathcal{M}}$ is a bijective function of $U$ that maps base coordinates on $\mathcal{M}$ to points on $\widetilde{\mathcal{M}}$, and  $J(U) = \det \frac{\partial \widetilde{U}}{\partial U}$ is the corresponding Jacobian.
A straightforward way to evaluate the second line of Eq.~\eqref{eq:def} is to sample configurations $U$ from the original probability measure $e^{-S(U)}/Z$ and instead evaluate the \emph{deformed observable}
\begin{equation}
    \mobs(U) \equiv e^{-\left[ \Seff(U) - S(U) \right]} \obs(\widetilde{U}(U)),
    \label{eq:obsTilde}
\end{equation}
where $\Seff(U) \equiv S(\widetilde{U}(U)) - \log J(U)$.
Cauchy's theorem guarantees an identical mean
\begin{equation} \label{eq:obsEqObsTilde}
    \langle \obs(U) \rangle = \langle \mobs(U) \rangle.
\end{equation}
Throughout this work, $\left< \cdot \right>$ denotes expectation with respect to the original probability density $e^{-S(U)} / Z$.
Since the distribution used for MC sampling is not modified in the deformed observable approach, 
integrals over many possible manifolds can be estimated using a single MC ensemble; this property is useful for both contour optimization and calculations with deformed observables.

This approach should be expected to work well unless the magnitude of the deformed observable fluctuates severely, which may occur if there is an overlap problem between $\Seff(U)$ and $S(U)$. In this case, one must sample from modified weights and use reweighting to compute Eq.~\eqref{eq:def}; this was done for path integrals with complex actions in Refs.~\cite{Alexandru:2015sua,Alexandru:2016gsd,Alexandru:2017czx,Alexandru:2017lqr,Tanizaki:2017yow,Alexandru:2018brw,Alexandru:2018ngw,Alexandru:2018fqp,Alexandru:2018ddf,Kashiwa:2018vxr,Kashiwa:2019lkv}. When many observables are needed, however, the cost of repeated MC ensemble generation based on each new manifold will be high. We therefore only consider contour deformations with good overlap between $\Seff(U)$ and $S(U)$ and apply the deformed-observable approach throughout this work.

\ssection{Optimizing the variance}
Though manifold deformations leave expectation values unchanged, they modify the variance of observables with complex phase fluctuations.  We restrict our investigation to observables with purely real expectation value,\footnote{The general case follows by applying the techniques discussed here to $\left< \Re \mobs \right>$ and analagous techniques to $\left< \Im \mobs \right>$.} where it is sufficient to consider
\begin{equation}
   \begin{split}
    \Var(\Re \mobs) &= \left< (\Re \mobs)^2 \right> - \left( \Re \left< \mobs \right> \right)^2.
   \end{split}\label{eq:var}
\end{equation}
While $\left( \Re \left< \mobs \right> \right)^2 = \left( \Re \left< \obs \right> \right)^2$ is unaffected by the choice of manifold, the variance is modified because $\left< (\Re \mobs)^2 \right>$ is not the integral of a holomorphic function. For each observable $\obs$, the task is then to find an optimized manifold for which $\Var(\Re \mobs) \ll \Var(\Re \obs)$. If this can be achieved, the StN ratio
\begin{equation}
   \begin{split}
      \text{StN}(\Re \mobs) \equiv \frac{|\Re \left< \obs \right>|}{\sqrt{\Var(\Re \mobs)}}
   \end{split}\label{eq:StNdef}
\end{equation}
will be improved.

The manifold minimizing $\Var(\Re \mobs)$ depends on the properties of the observable, and there is no single contour deformation which optimizes the StN of all observables. To account for this non-uniqueness, we use the methods of Refs.~\cite{Alexandru:2018fqp,Alexandru:2018ddf}, minimizing the variance for each observable over a family of manifolds smoothly parameterized by a vector of real numbers $\mparam$. The choice of manifold $\widetilde{\mathcal{M}}(\mparam)$, defined by the map $\widetilde{U}(U; \mparam)$, can be numerically optimized using stochastic gradient descent based on MC estimates of
\begin{equation}
\begin{aligned} \label{eq:gradVar}
    & \grad_{\mparam} \Var(\Re \mobs)
    = \left< \grad_{\mparam} (\Re \mobs)^2 \right>
    = 2 \left< \Re \mobs \Re \grad_{\mparam} \mobs \right> \\
    &= 2 \left< (\Re \mobs) \Re \left( 
    \mobs\left[
    -\grad_{\mparam}\Seff + \frac{\grad_{\mparam} \obs(\widetilde{U})}{\obs(\widetilde{U})}
    \right]
    \right) \right>.
\end{aligned}
\end{equation}
Crucially, the manifold parameters can be iteratively improved without generating new ensembles. This technique is used to optimize the integration contour in the second example below.

\begin{figure}
    \centering
    \includegraphics{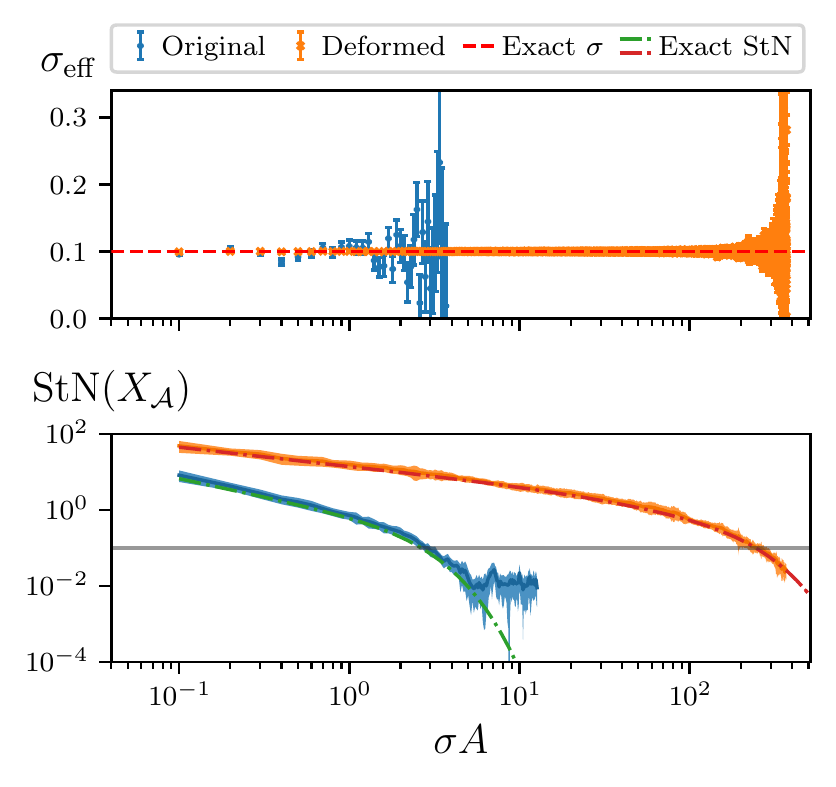}
    \caption{$\sigma_{\mathrm{eff}}$ and $\text{StN}(X_{\mathcal{A}})$ measured using the original integration contour ($\delta = 0$) in blue and the deformed integration contour ($\delta = 0.2$) in orange on an ensemble consisting of 10,000 MC samples with $\beta = 5.555$ and $L = 64$. The exact string tension $\sigma = 0.1$ is indicated by the dashed line (upper) and the exact StN scaling of Eq.~\eqref{eq:StNRW} by the dot-dashed lines (lower). The results for the original contour are truncated where the data become unreliable, at $A = L^2/32$.}
    \label{fig:string_tension_EMP}
\end{figure}

\ssection{Abelian gauge theory}
We first demonstrate the method on a two-dimensional $U(1)$ gauge theory with open boundary conditions. The central values and variances of observables can be computed analytically in this theory and are used to validate numerical results. The Wilson action~\cite{Wilson:1974sk} for $U(1)$ gauge theory in $2D$ can be expressed in terms of the plaquette $P_x = U^1_x U^2_{x+\hat{1}} (U^1_{x+\hat{2}})^\dagger (U^2_{x})^\dagger$ where $U^\mu_x \in U(1)$ and a square lattice $x_\mu \in \{ 0,\ldots, L \}$ for $\mu \in \{ 1,2 \}$ is used. Defining $\theta_x \equiv \arg P_x$, the action is given by
\begin{equation}
   \begin{split}
      S_G(\theta) \equiv -\beta \sum_x \cos \theta_x,
   \end{split}\label{eq:SP}
\end{equation}
where the sum excludes the sites on the open boundaries, i.e.~$x_\mu \neq L$.
In this theory, there is a change of variables with unit Jacobian to $\theta_x$ and residual degrees of freedom that can be trivially integrated out, allowing the partition function to be analytically evaluated as
\begin{equation}
   \begin{split}
      Z &\equiv \int \prod_x \left[ \frac{d \theta_x}{2\pi} e^{\beta \cos \theta_x } \right] = I_0(\beta)^{V},
   \end{split}\label{eq:Z}
\end{equation}
where $I_n(\beta)$ is a modified Bessel function and $V\equiv L^2$.
Expectation values of Wilson loops in this theory follow area law scaling
\begin{equation}
   \begin{split}
      \left< W_\mathcal{A} \right> &\equiv \left< \prod_{x\in\mathcal{A}} e^{i\theta_x} \right> = e^{-\sigma A},
   \end{split}\label{eq:Warea}
\end{equation}
where $A$ is the area of the region $\mathcal{A}$ enclosed by the loop in lattice units, and the string tension $\sigma$ is given by
\begin{equation}
   \begin{split}
      \sigma = \ln \left[ \frac{I_0(\beta)}{I_1(\beta)} \right].
   \end{split}\label{eq:sigma}
\end{equation}
In  Monte Carlo calculations, $W_{\mathcal{A}}$ has an exponential StN problem:
\begin{equation}
   \begin{split}
      \text{StN}( W_\mathcal{A} ) = \frac{ e^{-\sigma A} }{ \sqrt{ \frac{1}{2} + \frac{1}{2} e^{-\sigma^\prime A } -  e^{-2\sigma A} } },
   \end{split}\label{eq:StN}
\end{equation}
where $\sigma^\prime = \ln\left[ I_0(\beta) / I_2(\beta) \right]$.

The Wilson loops $W_{\mathcal{A}}$ can also be evaluated using the deformed observables approach. We consider manifolds defined by deformed variables
\begin{equation}
   \begin{split}
      \widetilde{\theta}_x = \theta_x + i \delta_x,
   \end{split}\label{eq:tildeTheta}
\end{equation}
where $\delta_x \in \mathbb{R}$ is a constant for each site $x$. This contour deformation has unit Jacobian and is smoothly connected to the original integration contour for any choice of $\delta_x$. Since the integrand $W_{\mathcal{A}} e^{-S_G}$ is holomorphic in $\theta_x$, the deformed observable gives unbiased estimates  of the expectation value $\left< W_{\mathcal{A}} \right>$. This is verified in Fig.~\ref{fig:string_tension_EMP}, where the analytically known string tension $\sigma^{\text{eff}} \equiv - \partial_A \ln W_\mathcal{A} = \sigma$ is reproduced by MC calculations on both the original and deformed integration manifolds.

Within this set of manifolds, we define a simpler one-parameter family by $\delta_x = \delta$ for $x \in \mathcal{A}$ and $\delta_x = 0$ otherwise. This parameterization is motivated by the limit of small phase fluctuations, valid at fine lattice spacing, in which the imaginary component of the action can be expanded for $\theta_x \ll 1$,
\begin{equation}
   \begin{split}
      \Im S_G(\widetilde{\theta}(\theta)) &= -\beta \sum_{x\in \mathcal{A}} \sin(\theta_x) \sinh(\delta) \\
      &= -\beta \sinh(\delta) \text{arg}(W_\mathcal{A}) [ 1 + O(\theta^2)].
   \end{split}\label{eq:interference}
\end{equation}
When $\delta$ is chosen such that $\beta \sinh(\delta) \approx 1$, $\Im S_G$ destructively interferes with the phase of $W_{\mathcal{A}}$. The manifold deformation simultaneously affects the magnitude of the deformed observable,
\begin{equation} \label{eq:Wtilde}
    \begin{split}
    X_{\mathcal{A}} &\equiv 
        J(\theta) e^{- \left[S_G(\widetilde{\theta}(\theta)) - S_G(\theta) \right]} W_{\mathcal{A}}(\widetilde{\theta}(\theta)) \\
    &= e^{- \left[S_G(\widetilde{\theta}(\theta)) - S_G(\theta) \right]} e^{-\delta A} W_{\mathcal{A}}(\theta),
    \end{split}
\end{equation}
heuristically replacing delicate cancellations of fluctuating phases with a reduced magnitude on each sample.

The StN effects of contour deformation can be more quantitatively understood by direct calculation from the path integral definition. One finds:
\begin{equation}
   \begin{split}
      \text{StN}( X_\mathcal{A} ) &= \frac{ e^{-\left( \sigma - \delta - \frac{1}{2}\sigma_\delta \right) A} }{ \sqrt{ \frac{1}{2} + \frac{1}{2} e^{-(\sigma_\delta^\prime-\sigma_\delta) A} - e^{-(2 \sigma - \sigma_\delta) A} } },
   \end{split}\label{eq:StNRW}
\end{equation}
where
\begin{equation}
  \begin{split}
      \sigma_\delta &= -\ln \left[ \frac{ I_0(\beta(2\cosh \delta - 1)) }{ I_0(\beta) } \right], \\
      \sigma_\delta^\prime &= -\ln \left[  \frac{ I_2(\beta\sqrt{5-4\cosh\delta}) }{I_0(\beta)} \left( \frac{e^{\delta} - \frac{1}{2}}{e^{-\delta} - \frac{1}{2}} \right)\right].
  \end{split}\label{eq:noisetension}
\end{equation}
Maximizing the StN as a function of $\delta$, the optimal integration contour defining $X_\mathcal{A}$ is found to have little sensitivity to $A$. For instance, at the finest gauge coupling used in this work ($\beta =5.555$, corresponding to $\sigma = 0.1$) the optimal $\delta$ is found to vary between $\delta \approx 0.204$, for $A = 1,$ to $\delta \approx 0.197$, for $A = 1000$. As shown in Fig.~\ref{fig:string_tension_EMP}, when $A \gg 1/\sigma$ the StN of $X_{\mathcal{A}}$ for a nearly optimal contour (where $\delta = 0.2$) is improved by orders of magnitude relative to the undeformed case. For example, when $A = 100 / \sigma$, the StN improves by a factor of $10^{43}$.

We have further confirmed that deformed observables are useful over a range of lattice spacings. Using ensembles of 10,000 samples each with lattice size $L = 64$, we investigate string tensions tuned to $\sigma = \left\{ 0.4, 0.3, 0.2, 0.1 \right\}$ in lattice units by fixing $\beta = \left\{ 1.843, 2.296, 3.124, 5.555 \right\}$. This corresponds to lattice spacing varying by a factor of two across the ensembles. By choosing a nearly optimal $\delta$ for every coupling, constant fits to $\sigma^{\mathrm{eff}}$ estimated from the deformed observable give results for $\sigma$ improved by $5\times$ -- $75\times$ in precision. The most benefit was found on the ensemble with finest lattice spacing ($\beta = 5.555$).

\begin{figure}
    \centering
    \includegraphics{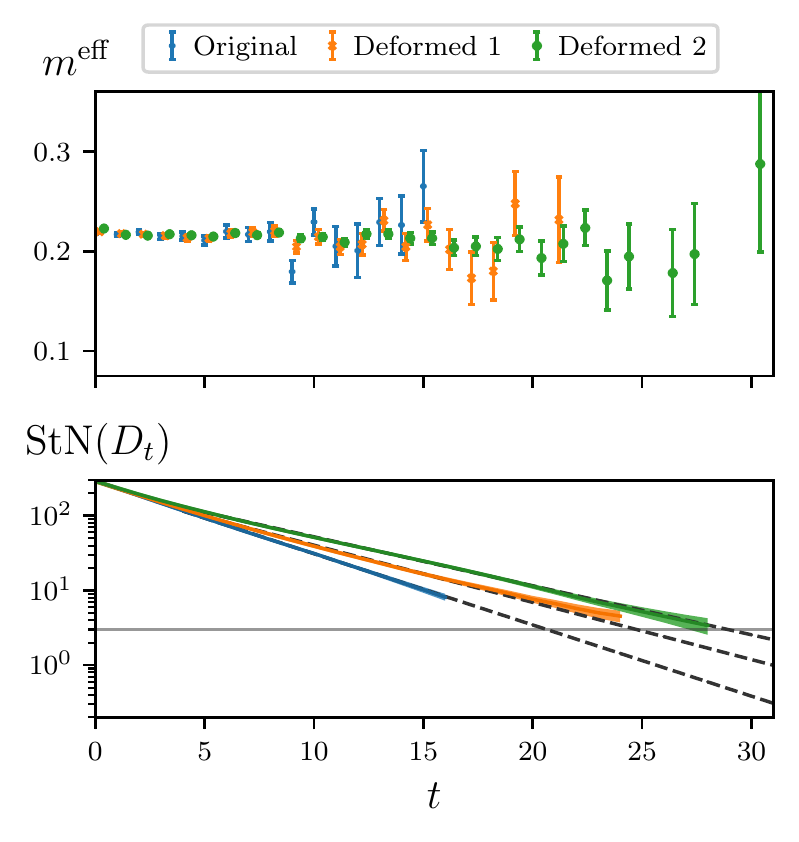}
    \caption{$m^{\mathrm{eff}}$ and $\text{StN}(D_t)$ in the complex scalar theory measured using the original integration contour (blue), a manually-tuned one-parameter contour (orange), and a numerically-optimized contour (green) on the ensemble consisting of 10,000 MC samples with the largest bare coupling considered here ($\lambda = 0.003$). Dashed exponential fits to the StN suggest growing improvement in the large-time limit where the data become unreliable at this finite ensemble size.}
    \label{fig:mass_EMP}
\end{figure}

\ssection{Complex scalar field theory}
To explore the generality of the deformed observables approach, we further apply it to complex scalar field theory in $0+1$D with a quartic interaction. Employing polar coordinates for the scalar field $\phi_t = R_t e^{i \theta_t}$, the lattice action reads
\begin{equation}
S=-2\sum_{t=0}^{L-1}{R_{t}R_{t+1}\text{cos}(\theta_{t+1} - \theta_{t})} + V(R),
\label{eq:lat-axn-phi4}
\end{equation}
where $V(R) = \sum_t{(2+m^2)R_t^2 + \lambda R_t^4}$, and periodic boundary conditions are used, $\theta_0 \equiv \theta_{L}$ and $R_0 \equiv R_{L}$.
Comparing this action with Eq.~\eqref{eq:SP}, it is apparent that phase differences $\theta_{t+1} - \theta_t$ in this theory have weights in the action with similar form to plaquettes in the $U(1)$ gauge theory. We therefore complexify the integration domain in a similar manner to the $U(1)$ case, deforming the phases as
\begin{equation}
\tilde \theta_t = \theta_t + i \delta^{(1)}_t  + i \delta^{(2)}_t f_c(R_t R_{t+1}) + i \delta^{(3)}_t f_c(R_{t-1} R_{t})
\label{eq:deformation}
\end{equation} 
while the $R_t$ remain undeformed. Here, $\delta_t^{(i)}$ are real parameters assigned to each lattice site and $f_c(x) = c \tanh (1 / c x)$ is chosen as a regularization of the function $1/x$ defined by a single additional parameter $c$. This form is motivated by an expansion in small phase fluctuations, while regularizing the function $1/x$ avoids overlap problems. Every manifold in the family defined by Eq.~\eqref{eq:deformation} has unit Jacobian, allowing efficient computation of the deformed observable.

The mass of the scalar particle is a key quantity in this theory and can be extracted from the large-time behavior of the single-particle propagator, $G_t \equiv \langle \phi_t \phi^{\dagger}_0\rangle$, using a local estimator $m^{\mathrm{eff}}(t) \equiv \arccosh\left( \frac{G_{t-1} + G_{t+1}}{2 G_{t}} \right)$. Written as a holomorphic function of the chosen variables,
\begin{equation} \label{eq:scalarGreensFn}
G_t = \left< R_t R_0 e^{i\theta_t - i \theta_0} \right> \equiv \left< C_t(R,\theta) \right>.
\end{equation}
At large times, $G_t$ has severe phase fluctuations and a StN problem arising from an exponentially falling signal and $O(1)$ variance  \cite{Detmold:2018eqd}.

We compare the original estimator based on direct evaluation of Eq.~\eqref{eq:scalarGreensFn} to the deformed observable defined by the manifold in Eq.~\eqref{eq:deformation},
\begin{equation}
\begin{aligned}
    D_t &\equiv e^{-\left[ \Seff(\theta) - S(\theta) \right]} C_t(R,\widetilde{\theta}) \\
    &= e^{-\left[ \Seff(\theta) - S(\theta) \right]} R_t R_0 e^{i \widetilde{\theta}_t - i \widetilde{\theta}_0}.
\end{aligned}
\end{equation}
We optimize the full $(3L+1)$-parameter form in Eq.~\eqref{eq:deformation} using the numerical approach based on gradient estimates defined by Eq.~\eqref{eq:gradVar}, and as a comparison optimize a simpler one-parameter subfamily of deformations defined by $c = \delta^{(2)}_{t'} = \delta^{(3)}_{t'} = 0$, $\delta^{(1)}_{t'} = t' \delta$ for $|t'| < t$, and $\delta^{(1)}_{t'} = t \delta$, which achieves destructive phase interference for small phase fluctuations.
Fig.~\ref{fig:mass_EMP} contrasts the results of the deformed observables to the original observable on a representative ensemble defined by bare $m^2 = (0.15)^2$, $\lambda = 3 \times 10^{-3}$, and $L = 64$. For any $t$, the statistical uncertainty on $m^{\mathrm{eff}}(t)$ is smaller on the deformed manifolds than on the original manifold. In comparison to the original manifold, the numerically optimized manifold reduces the observed exponential rate of StN degradation by $32\%$, while one-parameter optimization gives a reduction of $18\%$.

We find that the method is robust across several choices of bare couplings, $\lambda = \left\{ 0, 1, 2, 3 \right\} \times 10^{-3}$, ranging from the free theory to values well outside the regime of lattice perturbation theory~\cite{Alexandru:2017lqr}. Fits to the mass of the scalar particle in the original and deformed contour approaches agree to within statistical errors on ensembles consisting of 10,000 samples generated using Hybrid Monte Carlo~\cite{Duane1987HMC}. Excited state effects are not significant in this toy model; however, excited-state contamination prevents reliable single-exponential fits to correlation functions $G_t$ in more complex theories such as lattice QCD at small separations $t$. Here, we consider constant fits to $m^{\text{eff}}$ for fit ranges beginning at $t_i = \{5, 10\}$ to investigate improvement due to the deformed observable at a range of $t$. All fits have acceptable $\chi^2/N_{\text{dof}}$, and the scalar particle mass is determined more precisely by the deformed observable than the original observable, e.g.~for $\lambda = 3 \times 10^{-3}$ the fit beginning at $t_i = 5$ results in an estimate of the scalar mass $M$ that is $2\times$ more precise [$M_{\text{deform}} = 0.2166(11)$ vs.~$M_{\text{orig}} = 0.2150(20)$] and the fit beginning at $t_i = 10$ results in an estimate that is $3\times$ more precise [$M_{\text{deform}} = 0.2155(20)$ vs.~$M_{\text{orig}} = 0.2212(63)$].

\ssection{Conclusions}
Fluctuations of the complex phases of path integrands lead to sign and signal-to-noise problems in Monte Carlo calculations, both for theories with complex actions and for observables with complex phase fluctuations. Deforming the integration contours of path integrals can reduce these phase fluctuations and improve StN ratios for observables while ensuring the correctness of the results obtained. By interpreting the integrand on the deformed contour as a modified observable and optimizing the choice of deformation, results with lower variance can be obtained without modifying MC ensemble generation or generating new samples.

In low-dimensional Abelian gauge theory and complex scalar field theory, simple contour deformations inspired by a small phase fluctuation expansion are seen to reduce the variance of large Wilson loops and large-time correlation functions by orders of magnitude, resulting in improved estimates of physical quantities. Multi-parameter deformations obtained by numerical optimization result in even greater variance reduction. The methods presented here are general and apply to more complicated theories such as QCD, where it is similarly possible to complexify the fields and make path integral contour deformations that change correlation function variance while leaving expectation values unchanged. However, high-dimensional numerical optimization of more complicated and expressive parameterizations may be needed to provide significant variance reduction in these theories.
Future studies will explore whether suitably optimized contour deformations can provide significant StN improvement for correlation function calculations in QCD and other complicated theories.

\begin{acknowledgments}
\ssection{Acknowledgments}
We thank Henry Lamm and Phiala Shanahan for helpful discussions. WD, GK, and MLW are supported in part by the U.S.~Department of Energy, Office of Science, Office of Nuclear Physics under grant Contract Number DE-SC0011090.
WD is also supported within the framework of the TMD Topical Collaboration of the U.S.~Department of Energy, Office of Science, Office of Nuclear Physics, and  by the SciDAC4 award DE-SC0018121.
NCW is supported by the U.S. DOE under Grant No. DE-FG02-00ER41132. 
MLW was supported in part by an MIT Pappalardo Fellowship.
This manuscript has been authored by Fermi Research Alliance, LLC under Contract No. DE-AC02-07CH11359 with the U.S. Department of Energy, Office of Science, Office of High Energy Physics.
\end{acknowledgments}

\bibliography{noise_refs}

\end{document}